\pdfoutput=1
\documentclass[journal]{IEEEtran}
\usepackage{cite}
\ifCLASSINFOpdf
\else
\fi
\usepackage{graphicx}
\usepackage{amsmath}
\usepackage{subfig}
\usepackage{float}
\usepackage{multirow}
\usepackage{textcomp}
\usepackage{xcolor}
\usepackage{textcomp}
\usepackage{siunitx}
\usepackage{textcomp}
\usepackage{lipsum}
\usepackage[flushleft]{threeparttable}
\usepackage{booktabs,caption}
\usepackage{dblfloatfix}

\begin{document}
\title{Understanding the Importance of Heart Sound Segmentation for Heart Anomaly Detection}


\author{Theekshana Dissanayake, Tharindu Fernando \IEEEmembership{Member, IEEE}, Simon Denman, \IEEEmembership{Member, IEEE}, Sridha Sridharan, \IEEEmembership{Life Senior Member, IEEE}, Houman Ghaemmaghami, Clinton Fookes,\IEEEmembership{Senior Member, IEEE} 

\thanks{T. Dissanayake, T. Fernando, S. Denman, S. Sridharan and C. Fookes are with the Speech
Audio Image and Video Technologies  (SAIVT) Research Lab, Queensland University of Technology, Australia. H. Ghaemmaghami is with the M3DICINE Pty Ltd.}}

\maketitle

\begin{abstract}

Traditionally, abnormal heart sound classification is framed as a three-stage process. The first stage involves segmenting the phonocardiogram to detect fundamental heart sounds; after which features are extracted and classification is performed. Some researchers in the field argue the segmentation step is an unwanted computational burden, whereas others embrace it as a prior step to feature extraction. When comparing accuracies achieved by studies that have segmented heart sounds before analysis with those who have overlooked that step, the question of\textit{ whether to segment heart sounds before feature extraction} is still open. In this study, we explicitly examine the importance of heart sound segmentation as a prior step for heart sound classification, and then seek to apply the obtained insights to propose a robust classifier for abnormal heart sound detection. Furthermore, recognizing the pressing need for explainable Artificial Intelligence (AI) models in the medical domain, we also unveil hidden representations learned by the classifier using model interpretation techniques. Experimental results demonstrate that the segmentation plays an essential role  in abnormal heart sound classification. Our new classifier is also shown to be robust, stable and most importantly, explainable, with an accuracy of almost 100\% on the widely used PhysioNet dataset.

\end{abstract}


\section{Introduction}
\label{sec:introduction}

\begin{figure}
    \centering
    \includegraphics[width=0.99\linewidth]{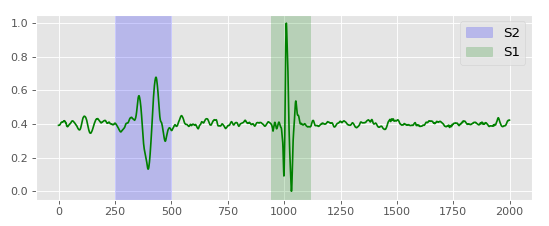}
    \caption{Phonocardiogram signal with S1 and S2 locations.}
    \label{pcg}
\end{figure}

Cardiovascular diseases have become one of the leading causes of death, and often lead to other medical conditions such as strokes, hypertension, heart failure and arrhythmia \cite{Santos2020OnlineOutlook,Patidar2014ClassificationTransform,Ghaemmaghami}. In the field of biomedical engineering, automatic abnormal heart sound detection can be considered a major prior step to cardiovascular disease diagnosis. The process of  identifying whether a given heart sound is normal or abnormal can be divided into three major steps: segmentation, feature extraction, and classification \cite{Clifford2017RecentAnalysis}. Firstly, the segmentation technique locates the fundamental heart sounds of the  Phonocardiogram (PCG) signal: S1 (first heart sound) and S2 (second heart sound); see Figure \ref{pcg}. However, detecting the fundamental heart sounds is itself a complex task, and can be affected by other internal sounds such as murmurs, the presence of third (S3) and fourth (S4) heart sounds and noise \cite{Zhang2019AbnormalSegmentation}. After segmenting the heart sound, various feature extraction techniques are used to extract features from the signal for training a classifier. These extracted features can generally be categorised as the time domain, frequency domain or time-frequency domain features of the PCG  wave. As the final step, using the extracted features, a classifier is developed to identify abnormal heart sounds. 

Studying these fundamental steps followed for normal-abnormal heart sound classification, the segmentation step can be considered an essential step to localize the signal before extracting various kinds of features. However several sophisticated studies in the literature, which have achieved superior performance for abnormal heart sound classification, have eschewed this step prior to feature extraction \cite{Zhang2019AbnormalSegmentation,Aroraheart,Sujadevi2019AnomalyLearning,Maknickas2017RecognitionCoefficientsb}. Therefore, whether segmentation is required prior to developing a classifier is still an open question that should be answered to determine if the additional computational burden of this process is beneficial.   

In summary, the principal objective of our research is to understand the importance of heart sound segmentation for the normal-abnormal heart sound classification task via a series of experiments involving empirical evaluations and model interpretation. Ultimately, the goal of this study is to develop a robust machine learning model for normal-abnormal heart sound classification while being able to explain the hidden representations learned by the model, and provide insight into the importance of segmentation for the overall task. The following list highlights the main contributions of our research: 

\begin{table*}[b!]
\centering
\renewcommand{\arraystretch}{1.1}
\caption{Recent advancements in heart anomaly detection achieved using the PhysioNet \cite{PhysioNet} database.}
\begin{tabular}{|p{3.6cm}|c|p{1.5cm}|p{5.3cm}|p{3.2cm}|p{0.8cm}|}
\hline
Study & Year & Segmented before? & Features & Model & Acc(\%) \\
\hline

Zhang~et~al.~\cite{Zhang2019AbnormalSegmentation}& 2019 & No & Temporal Quasi-Periodic Features & LSTM & \textbf{94.66}  \\

Numan~et~al.~\cite{jbhi} & 2019 & Yes & MFCCs, Time-Frequency &  Markov-Switching Model & 91.20 \\

Chowdhury~et~al.~\cite{Chowdhury2019Real-TimeMonitoring} & 2019 & Yes & Statistical, Entropy, Energy-based Features & Ensemble (Tree) & 94.63  \\ 

Arora~et~al.~\cite{Aroraheart} & 2019 & No & Statistical, Frequency Domain, MFCCs & XGBoost & 92.90  \\

Alaskar~et~al.~\cite{Alaskar2019TheClassification} & 2019 & Yes & AlexNet Model-Extracted & SVM & 87.00 \\

Sujadevi~et~al.~\cite{Sujadevi2019AnomalyLearning} & 2018 & No & Raw Signal & RNNs, CNN &79.80  \\

Kay~\&~Agarwal~\cite{Kay2017DropConnectedSounds}~& 2017 & Yes & Time-Frequency, Inner Beat Features & Neural~Network & 85.20 \\

Maknickas~\&~Maknickas~\cite{Maknickas2017RecognitionCoefficientsb} & 2017 & No & MFCCs & CNN & 84.15  \\

Whitaker~et~al.~\cite{Whitaker2017CombiningClassification} & 2017 & Yes & Time, Frequency, Sparse Coding & SVM & 80.30 \\

Potes~et~al.~\cite{ense} & 2016 & Yes & PCG  Amplitudes, Power, Frequency-based & ADABoost + CNN & 94.24  \\ 

\hline
\end{tabular}

\begin{tablenotes}
      \small
      \item \textbf{Acc(\%)}: Accuracy,\textbf{MFCCs}: Mel-Frequency Cepstral Coefficients, \textbf{LSTM}: Long-Short Term Memory neural networks, \textbf{RNN}: Recurrent Neural Networks, \textbf{CNN}: Convolution Neural Networks, \textbf{SVM}: Support Vector Machine, \textbf{MLP}: Multi-Layer Perception, 
\end{tablenotes}

\label{abnormal}
\end{table*}

\begin{enumerate}

\item We empirically evaluate the importance of heart sound segmentation as a prior step to heart sound classification, and then, propose a novel deep learning model which achieves an accuracy of 98.71\% for heart sound classification on the PhysioNet \cite{PhysioNet} dataset. 

\item Going beyond the quantitative results, we interpret the developed deep learning model using the SHAP (SHapley Additive exPlanations \cite{LundbergAPredictions}) algorithm to reveal the hidden representations learned by the model.

\item Based on insights obtained from the first experiment regarding segmentation, a second architecture (a slight variant) which achieves close to 100\% accuracy is proposed.

\item We propose a procedure for interpreting models with temporal data, especially signals, using the SHAP algorithm and Occlusion maps; allowing us to critically evaluate the role of segmentation.

\end{enumerate}

\section{Related Work}

\subsection{Heart Sound Anomaly Detection}

Table \ref{abnormal} outlines recent studies on abnormal heart sound detection summarising the employed methods, features and prediction accuracies gained by the respective techniques. Along with these measures and design methodologies, this table shows whether the considered study has used segmentation prior to feature extraction or not. It should be noted that this table only represents the most recent studies on normal-abnormal heart signal classification which demonstrate state of art performance.

Reviewing the accuracies gained for heart anomaly detection, both Ensemble Learners and various classes of Neural Networks demonstrate better performance compared to traditional machine learning algorithms. However, it should be noted that most best-performing methods in the literature have employed pre-extracted features from the PCG wave, and the results achieved so far show promising evidence for PCG signal based medical diagnosis.

From consideration of the studies in Table \ref{abnormal}, the answer to the question of \textit{whether to segment heart-sounds before performing classification} remains unresolved. This concern was also raised by  Clifford~et~al.~\cite{Clifford2017RecentAnalysis} in their extensive review of heart sound analysis research from the years 2015 to 2017. Moreover, according to the recent study by Zhang~et~al.~\cite{Zhang2019AbnormalSegmentation}, which introduces the best-performing model to date, segmentation is not a mandatory step for heart anomaly detection because the primary focus of the algorithm should be detecting the presence of an anomaly, which can be achieved without locating it. However, examining the recent literature, studies that considered segmentation as a prior step have also achieved competitive performance \cite{Chowdhury2019Real-TimeMonitoring,ense}. Therefore further investigations  are required.

In addition to understanding the importance of segmentation for PCG signal-based classification tasks, another primary concern when applying machine learning  to the medical domain is the trust associated with  predictions made by a model. Even when a model performs with excellent accuracy, unless its behavior and predictions can  be explained, a medical expert or a patient may not trust the validity of the system \cite{Holzinger2017WhatDomain, Adadi2018PeekingXAI}. Hence, there is an essential need to advance machine learning-based medical diagnosis research to a state where these concerns can be effectively addressed. 

It is well established that, unlike traditional machine learning algorithms, deep learning models are better able to model complex relations in the data and thus achieve state-of-art results. However, these models have a black-box nature by virtue of their large number of parameters and  complex architecture. Even if the model shows good performance for a particular task, understanding what leads to the achieved performance can be complex \cite{Ribeiro2016WhyClassifier}. Fortunately, there are novel techniques that can be used to understand the nature of the model up to some extent (for eg: LIME \cite{Ribeiro2016WhyClassifier}, DeepLIFT \cite{ShrikumarLearningDifferences} and SHAP \cite{LundbergAPredictions}). 

Despite these advances in model understanding approaches, they have not been applied to bio-signal models, even though there is a pressing need for explainable AI within the medical domain. Motivated by this need, we interpret the designed models to discover which segments in the signal leads to the achieved performance. More importantly, this can be used to determine which segments (or segment combinations) in the heart sound indicate the presence of an anomaly.

\subsection{Model Interpretation}

\subsubsection{Shapley Values}

One way of understanding the hidden representations learned by a black-box model is examining how a particular feature in the input contributes to a decision made by the model. In the context of a simple linear classifier, $f(x)$, a prediction $\hat{f}(x)$ made for the $n$ dimensional instance $x$ can be expressed as,

\begin{equation}
\label{liner}
\hat{f}(x) = \beta_0 + \beta_{1}x_{1} + \beta_{2}x_{2} + \dots + \beta_{n}x_{n}.
\end{equation}

Here, each $x_j$ is the value of each feature and $\beta_{j}$ is the weight associated with each feature. 

The contribution factor, or in this case learned weights of a particular feature, might have a positive value or a negative value indicating the overall influence on the final prediction. Accordingly, if the contribution value is near zero or low compared to other absolute weights, it implies that the considered feature does not have a substantial contribution to the final prediction made by the model. Although this simple analogy explains a prediction made by a model with respect to the contributions made by the input features, this approach can not be adopted to explain the performance of the entire machine learning algorithm when applied to a particular application \cite{molnar2019}. 

Fortunately, the SHapley Additive exPlanations (SHAP) \cite{LundbergAPredictions} algorithm, designed by adopting Shapley values introduced in Game Theory \cite{shapley}, has the ability to provide such instance-level explanations for any machine learning model. According to this strategy, if a feature in the input is treated as a \textit{player}, and the final prediction made by the model is the \textit{playout}, then the computed Shapley values indicate how the playout is distributed among players. More precisely, the Shapley value of a particular feature is the contribution made by that feature to the final decision made by the model. 

The SHAP algorithm, an advanced version of Shapley values, is able to produce local and global explanations for a model. Local explanations provided by the SHAP algorithm imply how much a particular feature contributes (or how important it is) to the decision made by the model; and global explanations present insights into the overall importance of a particular feature for the predictions made by the model. It should be noted that, as with weights in a simple linear classifier, Shapley values can have both negative and positive signs. 

\subsubsection{Occlusion Maps}

Another way of understanding how input features or, feature regions impact predictions is by visualizing Occlusion Maps \cite{Zeiler2014VisualizingNetworks}. These maps provide insights into important feature regions that the model concentrates on while making a correct prediction. Simply put, this map is generated by masking the input feature map using a kernel and visualizing the activations (or probabilities) for the class of interest. For instance, if the masked region contains features that are important for making a correct prediction, predictions made by the model for the masked feature map will have high probabilities for the correct class. Similarly, if the masked region is not important, or in other words the model does not focus on that region to make a correct prediction, then, the map will have lower probabilities for the class of interest.  

Like Shapley values, this method can be seen as another way of understanding what regions in the input feature map contribute to a correct prediction. Unlike Shapley values, this technique can be seen as a local explanation (or an instance-level)  method. Furthermore, local Shapley values provide insights into the contribution of a particular feature for the prediction made. Sometimes, highly contributing features (or important features) are not present in the input, yet the model might be able to produce a correct prediction using the available features. Therefore, visualizing Occlusion maps provides additional insights into the importance of having a particular feature region in the input feature map. An occlusion map will also show whether the model is focusing on the expected region or is focusing on seemingly unimportant sections in the input (i.e background textures, noise). 

The rest of the paper is organized as follows. The next section of the study explains the data preparation and evaluation method adopted from \cite{Zhang2019AbnormalSegmentation}. Then, the next two sections explain the two experiments conducted, where the first experiment (Section \ref{exp1}) aims to understand the importance of segmentation; and the second experiment (Section \ref{exp2}) proposes a robust model by considering the observations made. Finally, Section \ref{disc} critically analyzes the results and implications. 

\section{Data Preparation and Evaluation}

The PhysioNet \cite{PhysioNet} database was selected as the main data source for investigation. This database contains an imbalanced dataset from clinical and non-clinical experiments and contains numerous variations including differences in device type, age and gender, variations in the  physical conditions of the subject while acquiring the data, and different device placement locations. In our evaluation we adopt the 10 Fold Cross-Validation procedure proposed by Zhang et al. \cite{Zhang2019AbnormalSegmentation} (i.e the best performing model in the literature).

Since in PhysioNet files are of different lengths, a windowing algorithm was employed to extract 1s signals from each wave file with a 0.1s shift (i.e 0.9s overlap). Recognizing the imbalanced nature of the PhysioNet database, studies in the literature have adopted various techniques to improve the model training process with imbalanced data. Unlike those, our investigation trains the model on a balanced database ($bal\_db$) which is not augmented or up-sampled. To create such database, all abnormal signals in the sample were selected. To ensure the the balanced database captured all normal wave files in the PhysioNet database, a sampling algorithm was employed. This algorithm selects windowed-signals from all normal wave files while keeping the balanced nature of the resulting data. Ultimately this process yields two databases, the balanced database ($bal\_db$) which is used to train/validate the model and the rest of the remaining (dropped) normal data ($rest$).

\begin{equation}
    \label{bal}
    bal\_db = [~[train_i,val_i]~for~i \in [1,2,\dots,10]~].
\end{equation}

As shown in Equation \ref{bal}, for each fold $i$, the model will be trained on the $train_i$ database, and the model will be evaluated on the $val_i$ database and the $rest$. Then, the ground truth signal level classification for each fold will be performed by majority voting. Ultimately, for each fold, the entire PhysioNet database will be employed, and the test set is used for each fold to compute accuracy, sensitivity and specificity as same as in previous studies  \cite{Zhang2019AbnormalSegmentation,Chowdhury2019Real-TimeMonitoring,Zabihiel}.

\section{Hybrid Models with Segmentation\label{exp1}}

This section describes the first experiment where we aim to determine the importance of segmentation using three deep learning architectures. We start the experiment with a quantitative evaluation concerning model accuracies to evaluate the performance which can be achieved by adopting segmentation. Following this, we use model interpretation to understand the hidden representations learned by the model. 

\subsection{Quantitative Evaluation}

Prior to investigating the importance of segmentation, a robust segmentation model that can be used for the experiments is needed. Hence, we use the segmentation model proposed by Fernando~et~al.~\cite{Fernando2019HeartAttention} which has an accuracy of 97\% using an Attention-based Long-Short Term Memory (LSTM) network. Since the considered architecture employs Mel-Frequency Cepstral Coefficients (MFCCs), and MFCCs have been extensively used in medical \cite{mendocafab}, speech \cite{mfccs} and emotion \cite{Swain2018DatabasesReview} domains, we also adopt MFCCs as the primary feature. 

Similar to \cite{Fernando2019HeartAttention}, we use six Mel-Frequency filter banks within the range of 30Hz-300Hz and corresponding Delta ($\Delta$) and Delta-Delta ($\Delta^2$) features of the MFCC spectrum, resulting in a feature map shaped $[6\times99\times3]$ for a 1s signal.

\begin{figure}[b!]
    \centering
    \includegraphics[width=\linewidth]{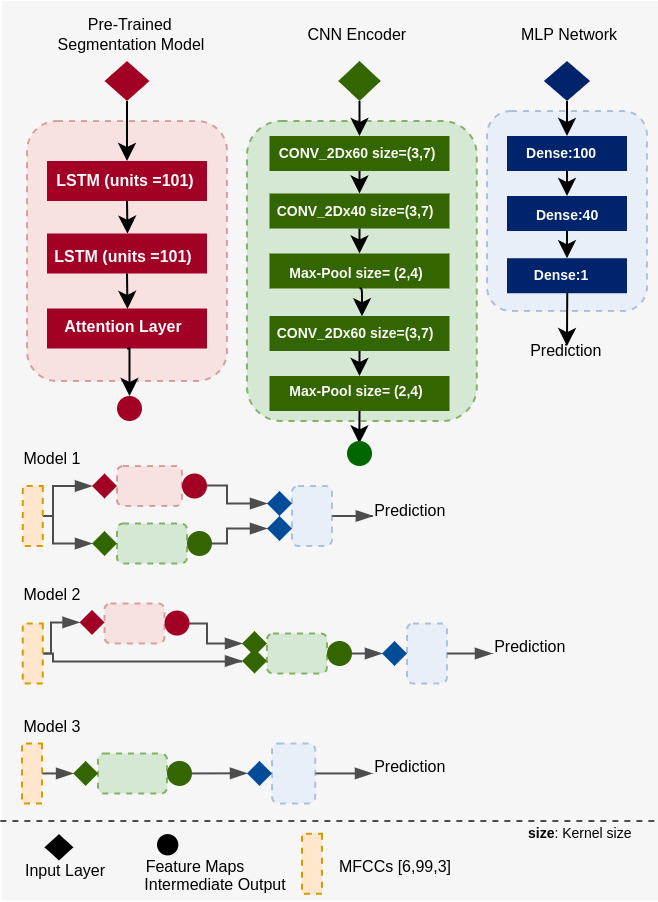}
    \caption{\textbf{Top:} Three deep network components used in the investigation. Left: The pre-trained segmentation model from \cite{Fernando2019HeartAttention}, Middle: The CNN encoder, Right: The MLP network with the last layer having a sigmoid activation. \textbf{Bottom Three:} Three deep learning architectures used in the experiment (Model 1,2,3).}
    \label{archs}
\end{figure}

Figure \ref{archs} presents the three main components employed in the experiment (top row) and the three proposed hybrid architectures derived by combining those components (bottom three). The first component is the pre-trained LSTM neural network adopted from \cite{Fernando2019HeartAttention} with the final layer removed. The second component implements a Convolution Neural Network (CNN) encoder to perform spatial feature learning on the supplied feature maps. 
The third component acts as the final function of the hybrid model. This is a simple Multi-layer Perseptron Network (MLP) with three hidden layers. This network accepts the learned representations from the previous feature extractors. Additionally, each layer of the MLP network has a 0.5 Dropout \cite{chollet2015keras} rate and a value of 3.0 for the MaxNorm Kernel Normalizer \cite{chollet2015keras}. Similarly, all convolution layers of the CNN network have values of  0.2 and 2.7 respectively for these hyper parameters. 

Three architectures are defined based on these components:

\begin{itemize}

\item \textbf{Model~1}: This architecture connects the segmentation model and the CNN encoder to the MLP network. The MLP network accepts a combined learned feature map of shape $[460\times1]$ by concatenating the features extracted by both models. 

\item \textbf{Model~2}: This model combines the CNN encoder and the MLP network while accepting the extracted segmentation-related feature map as an additional input channel. This feature map is generated by staking the transpose of the segmentor's output (shaped $[99\times1]$) vertically desired shape of $[6\times99]$, at which point the feature map can be used as an additional channel alongside the MFCC, $\Delta$ and $\Delta^2$ features; resulting in a final input shape of $[6\times99\times4]$~(*\textit{It should be noted that the transpose operation preserves the temporal relationships among the MFCCs and LSTM's outputs.})

\item \textbf{Model~3}: A CNN+MLP model without the feed from the segmentation model outputs (i.e the model without segmentation). 

\end{itemize}

\begin{table}[b!]
    \centering
    \begin{tabular}{|p{1.2cm}|p{1.8cm}|p{1.8cm}|p{1.8cm}|}
    \hline
    Model &  Accuracy(\%) & Sensitivity(\%) & Specificity(\%) \\ \hline
    
    Model~1  & 98.71($\pm$0.46) & 98.46($\pm$0.21) & 99.25($\pm$0.23)  \\
    Model~1*  & 98.49($\pm$0.13) & 98.19($\pm$0.28) & 99.36($\pm$0.43)  \\ \hline
    Model~2  & 84.95($\pm$0.36) & 92.50($\pm$0.34) & 70.57($\pm$0.27)  \\
    Model~2*  & 84.19($\pm$0.13) & 94.12($\pm$0.38) & 72.16($\pm$0.23)   \\ \hline
    Model~3  & 99.21($\pm$0.28) & 99.10($\pm$0.36) & 99.67($\pm$0.33)  \\
    Model~3*  & 98.94($\pm$0.27) & 98.81($\pm$0.24) & 99.46($\pm$0.21)  \\
    
    \hline
    \end{tabular}
    \caption{10 Fold Cross-Validation results for ground truth signals by considering majority voting. Model names end with a * only accept an MFCC feature map as the input.}
    \label{exp1}
\end{table}

\begin{figure*}[b!]
    \centering
    \includegraphics[width=0.99\linewidth]{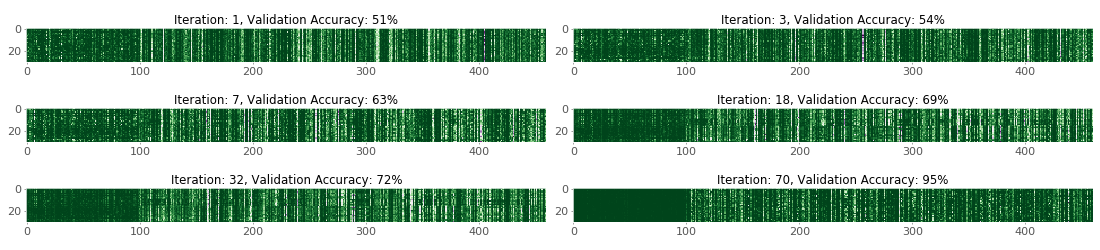}
    \caption{Shapley value variation while training the model. Shapley values with higher contributions to the output are indicated in white while dark-green indicates a low contribution. Each graph shows thirty absolute Shapley values vectors (shaped [$460\times1$])  of the intermediate layer for thirty randomly picked test instances (15 normal signals and  15 abnormal signals). The first 100 values represent the importance/contribution of features from the segmenter for the final prediction.}
    \label{accshap}
\end{figure*}

\begin{figure*}[b!]
    \centering
    \includegraphics[width=\linewidth]{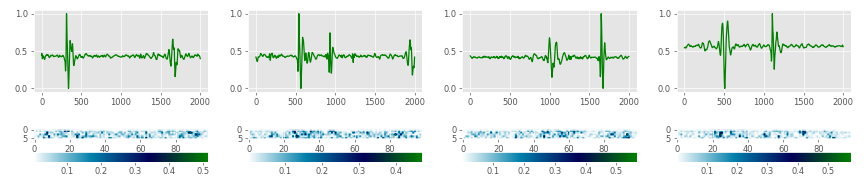}
    \caption{Absolute Shapely values for the input MFCC feature map $[6\times99\times1]$ showing a randomly selected ground truth normal signal tested on Model 1. High contributions are indicated by a dark blue-green contours in the feature maps, and are white otherwise (see the colour map scale).}
    \label{shapin}
\end{figure*}

\begin{figure*}[t!]
    \centering
    \includegraphics[width=\linewidth]{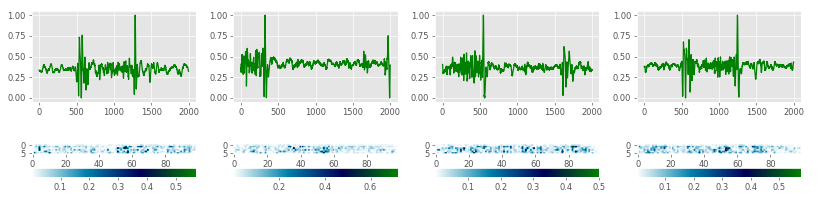}
    \caption{Absolute Shapely values for the input MFCC feature map $[6\times99\times1]$ showing a randomly selected ground truth abnormal signal tested on Model 1. High contributions are indicated by a dark blue-green contours in the feature maps, and are white otherwise (see the colour map scale).}
    \label{shapinab}
\end{figure*}

Table \ref{exp1} presents the 10 Fold Cross-Validation results of the proposed model architectures. Additionally, this table also shows three variants of the model (Model $i^*$) that only use the MFCC feature map as the input, though they do provide the full feature map to the segmentation model (i.e MFCC, $\Delta$ and $\Delta^2$)  if that component is used. By examining the results, models accepting MFCC feature maps tend to perform with similar accuracy to models recieving all three feature maps. Another interesting fact that can be observed is the model combined with the intermediate segmentation feature map (Model 1) has similar accuracy compared to the model without segmentation (Model 3).  However, the model which accepts the segmentation feature map as an input to the CNN encoder (Model 2) has a significantly lower prediction accuracy. Collectively, the presented results introduce a set of robust and stable classifiers that outperform the best performing model in the literature, Zhang et al.~\cite{Zhang2019AbnormalSegmentation} by almost 4\%.

According to the results, feeding the learned representation from the segmentor seems to introduce some additional complexity or noise to the input feature map. Furthermore, models receiving an MFCC feature map produce similar results to the models which receive multiple feature maps. This might be explained by $\Delta$ and $\Delta^2$ features being slightly different versions of the MFCC map, which could be derived through a simple convolution kernel (if necessary). 

Examining all results, it is apparent that the segmentation step does not significantly improve the model prediction accuracy, however it is important to try to understand why segmentation appears to offer limited or no benefit.

\subsection{Model Interpretation}

We first analyse the intermediate layer of Model 1 where the segmentation information is supplied to the MLP network. The dimension of this layer is $[460\times1]$ where the first $[100\times1]$ features are from the segmentation model. Figure \ref{accshap} shows absolute Shapley values computed for 15 normal and 15 abnormal testing instances (signal feature maps), while the model holding different validation accuracies (i.e as the model trains). These absolute Shapley values represent the contribution of a particular intermediate feature to the final output. If the individual contribution is high, they are clearly visible as white contours in the feature maps, and low contributing features are dark green. The first 100 values are the features from the segmenter. Examining the heat maps, it is apparent that the model is  learning to focus on the CNN encoder features instead of segmentation-based features, leading CNN-based features to have a higher contribution to the final prediction. This factor seems to be valid for all considered testing instances. Therefore, according to this insight, even with the segmentation feature map present, the model is focusing on CNN features. However, according to Figure \ref{shapin}, this conclusion might not be entirely valid regarding the importance of segmentation to the final result.

Figure \ref{shapin} is generated from absolute Shapley values of four normal signal feature maps from the dataset. The figure only demonstrates Shapley value maps of the first feature dimension (i.e the MFCC map $[6\times99\times1]$). Higher values within the Shapley maps can be seen in the S1 and S2 locations, suggesting that this area of the signal has a greater contribution when making a decision. This is visible in almost all generated Shapley value explanations from a randomly picked sample of size 60. It should be noted that, unlike $\Delta$ and $\Delta^2$ maps, MFCCs have a strong temporal relationship with the ground truth signal, and therefore, aligning them with the temporal axis of the ground truth signal helps to understand the model. 

By examining the Shapley value maps of the sample, S1 location-based MFCCs seem to be more important for prediction than S2-based MFCCs. Therefore, seemingly, the model has the ability to focus on S1 or S2 locations of the PCG wave without using the additional knowledge provided by the segmentation model (at the beginning or intermediate level).  Unfortunately, as shown in Figure \ref{shapinab}, the same algorithm applied to abnormal signals does not provide such interpretable results. By considering these observations, it is apparent that without utilizing the knowledge provided by the segmentation model, the CNN+MLP network is capable of learning a complex function that can distinguish between normal-abnormal sounds while being able to localize the signal if needed.

In summary, these results highlight three main points regarding the segmentation of heart sounds and normal-abnormal heart sound classification. 

\begin{itemize}

\item Considering the classification accuracy of three models and their variants, these models perform well even with a single MFCC feature map. In fact, the accuracy difference between having MFCCs only compared to MFCCs, $\Delta$ and $\Delta^2$ is negligible. 

\item Regarding the importance of segmentation, if the model is powerful enough, it will be able to learn to segment the data. Furthermore, Shapley values indicate that the MFCC features near the S1 and S2 locations have a higher contribution to prediction.  Therefore, segmentation can be seen as an essential requirement for the classifier to make accurate predictions. 

\item Feeding the knowledge separately may not generate the features that the model desires. One reason for this may be the model is learning a pattern that has a strong association with the S1 and S2 locations, which can only be extracted by the model itself.

\end{itemize}

Examining these conclusions, it is apparent that the MFCCs and the structure of the proposed architecture plays an important role in the performance of the classifier. Given the above findings, in the next experiment, we propose a variant of the same model architecture with an enhanced MFCC feature map.

\section{A robust model without segmentation\label{exp2}}
\begin{figure}[b!]
    \centering
    \includegraphics[width=0.65\linewidth]{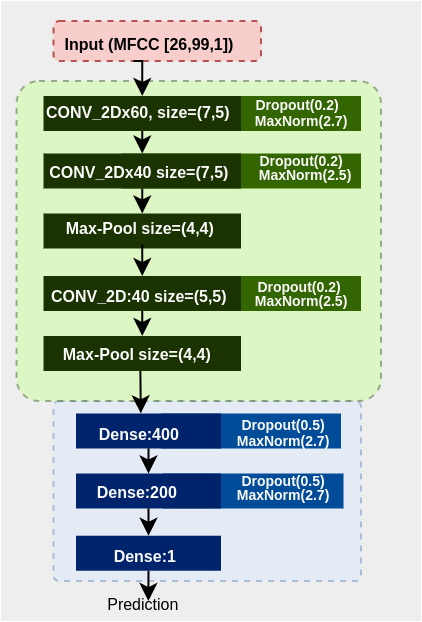}
    \caption{Proposed Architecture with three Convolution layers and three Dense layers. The input to the network is an MFCC feature map extracted from a 1s signal shaped $[26\times99\times1]$}
    \label{finaml}
\end{figure}

\begin{figure*}[b!]
    \centering
    \includegraphics[width=0.99\linewidth]{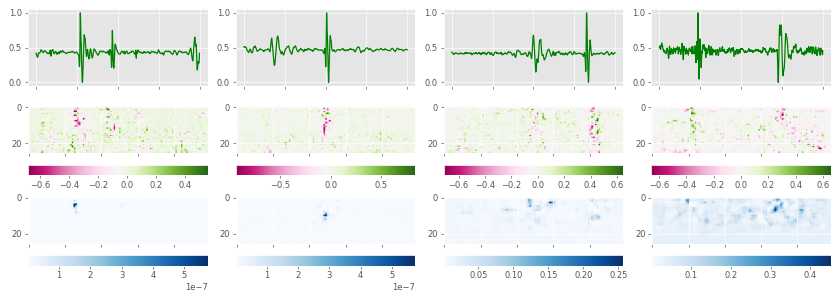}
    \caption{Shapley values (middle) and Occlusion maps (bottom) for four MFCC feature maps with a normal ground truth signal (top, four signals from a randomly picked sample of 60, shaped: $[26\times99\times1]$). Shapley Maps shows the Shapley values with the sign (negative: pink, positive: green), Occlusion maps shows the classifier output if the input feature region is masked using a $[3\times3]$ kernel.}
    \label{finalnormal}
\end{figure*}

\subsection{Model Architecture}

Figure \ref{finaml} shows a detailed diagram of the proposed architecture. The new  architecture has the same organization as the Model 3 proposed in the Section \ref{exp1}. However, the MFCC feature map used in the proposed model is computed from 26 Mel-frequency filter banks within an extended frequency range of 0-500Hz. In contrast with the previous structure, this model uses wider convolution filters due to the new expanded feature map. Furthermore, since the resulting feature map  from the CNN encoder holds a higher number of intermediate features (shaped $[3\times6\times60]$), the MLP network of this architecture has been also modified. 

The proposed model achieves  99.78~($\pm$0.22)\% accuracy for abnormal heart sound classification. Furthermore, the model achieves 99.77~($\pm$0.12)\% sensitivity and  99.72~($\pm$0.18)\% specificity for detecting abnormal heart sounds. The designed model outperforms the best performing model in the literature, Zhang et al. \cite{Zhang2019AbnormalSegmentation}, with an accuracy gain of almost 5\%. Furthermore, the proposed classifier outperforms models by Chowdhury et al. \cite{Chowdhury2019Real-TimeMonitoring} and Zabihi et al. \cite{Zabihiel} which adopts the same analysis strategy. 

As an additional step, we also evaluate the model on the murmur database created by Yaseen~et~al.~ \cite{Yaseen2018ClassificationFeatures}. This database contains 200 normal heart sounds and 800 abnormal murmur sounds (200 each for Aortic Stenosis, Mitral Regurgitation, Mitral Stenosis and Mitral Valve Prolapse). Following the evaluation protocol of \cite{Zhang2019AbnormalSegmentation}, we achieve performance of  99.97~($\pm$0.23)\% accuracy, 99.98~($\pm$0.31)\% sensitivity and 99.96~($\pm$0.26)\% specificity.

\subsection{Computational Efficiency and Stability}

The standard deviation of the 10-Fold Cross-Validation result shows that the designed model is consistent and converges to almost the same classification accuracy for every fold. The model only takes 3s to make 20000 predictions on a computer with 12GB memory, six Intel E5-2680 2.50 GHz CPUs and one Nvidia K40 GPU.  Hence, the proposed model represents a simple,  stable and efficient architecture for abnormal heart sound classification.

\subsection{Model Interpretation}

\begin{figure*}[t!]
    \centering
    \includegraphics[width=0.99\linewidth]{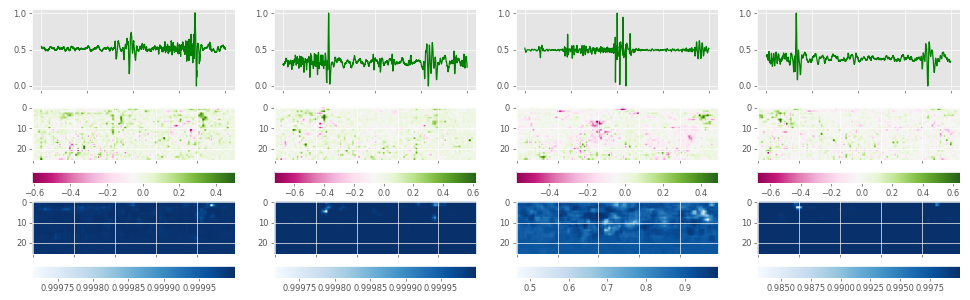}
    \caption{Shapley values (middle) and Occlusion maps (bottom) for four MFCC feature maps with a abnormal ground truth signal (top, four signals from a randomly picked sample of 60, shaped: $[26\times99\times1]$). Shapley Maps shows the Shapley values with the sign (negative: pink, positive: green), Occlusion maps shows the classifier output if the input feature region is masked using a $[3\times3]$ kernel.}
    \label{finalanormal}
\end{figure*}

The next step of the experiment strives to find the hidden representations learned by the classifier using Shapley values. As observed in the previous analysis, the most imporant feature for the predictions are near to the fundamental heart sound locations. Since Shapley values only provide insights related to the contribution of a particular feature for the final prediction, this research also employs Occlusion maps \cite{Zeiler2014VisualizingNetworks} (computed using $[3\times3]$ mask), which can be used to find the importance of a particular feature region by masking it from the input. To be more precise, the generated output presents the variation of the prediction if the considered feature region is not visible at the input.  

A set of randomly picked normal and abnormal signals were used to compute Shapley value maps and Occlusion maps for model interpretation (each sample had 60 signal files for the interpretation). As mentioned, this model uses 26 filter banks to generate the MFCC feature map resulting in an input shape of $[26\times99\times1]$. Figure \ref{finalnormal} and \ref{finalanormal} show Shapley values (middle) and Occlusion maps (bottom) for four ground truth normal and abnormal signals (top). The Shapley value graphs also illustrate whether the considered feature contributes negatively (pink) or positively (green) to the prediction made. The color map of the Occlusion map shows the output of the classifier if the considered region is masked (the decision boundary considered for the abnormal classification is 0.9).

Considering the patterns observed in the Shapley values and ground truth signals, almost all normal signal SHAP explanations show a pattern where the classifier is learned to focus on S1 and S2 locations within the PCG wave while making correct predictions. As mentioned, the previous Shapley computations were unable to provide such insights into what sort of signal locations contribute to correct abnormal signal prediction. However, Figure \ref{finalanormal} clearly shows that the model is concentrating on either S1 or S2 regions in the MFCC map.

As discussed, unlike Shapley values, Occlusion maps provide insights about the impact on the presence of a feature region in the input. By observing consecutive Shapley values and Occlusion maps, the features which are contributing most can be seen as the most important in the input. However, even without those features, the model might be able make correct predictions. This is because the other features that are aligned to S1 or S2 locations contribute to the decision (i.e Shapley values that have medium contribution factors). Furthermore, the same phenomenon is also visible in Figure \ref{finalanormal} where the Occlusion maps are computed for abnormal heart sounds. When examining the Occlusion map values for normal prediction, even the S1 location features are masked, the model is able to make correct predictions. Fortunately, this factor also seem to be valid for abnormal cases (i.e the computed probabilities/predictions are greater than 0.9, see Figure \ref{finalanormal}). Therefore, the model is robust enough to make correct predictions even if the part of the desired feature region is masked.

Collectively it is apparent that compared to the previous models, the model designed in Experiment II has the ability to locate S1 and S2 locations accurately. This is likely the reason for the model having a prediction accuracy of almost 100\%. Therefore, this result further enhances the outcomes gained in the previous experiment, and the assumptions made from those interpretations are confirmed.

\section{Conclusion\label{disc}}

The main objective of this research was to discover the advantages of heart sound segmentation for abnormal heart sound classification, and derive a robust explainable model using those insights.

Considering the models proposed in this study, Model 3 from Experiment I and the model described in Experiment II, have superior and stable performance for normal-abnormal heart sound classification compared to the state-of-the-art. The final model, which is designed by considering the conclusions made in the first experiment, outperforms the best performing classifier in the literature with an accuracy by 5\%. Furthermore, the architecture of the classifier only has three convolution layers and three fully connected layers, making it a  simple but efficient model.

Regarding the importance of segmentation, if the model has the capacity to learn the segmentation-function while extracting associated features from the S1 and S2 locations, then the segmentation is not needed as a prior step. This factor entirely depends on the robustness of the model, though as demonstrated in this work MFCC features do allow the identification of S1 and S2 locations. Ultimately, in conclusion, segmentation plays an essential role in abnormal heart sound detection. Another significant insight is, models from the investigations that overlooked segmentation \cite{Zhang2019AbnormalSegmentation,Aroraheart} might possess such capability within, but those insights were not evident or have not been investigated. Therefore, being the first investigation that uses model interpretation to understand abnormal heart sound classification, our study provide valuable insights into the compelling need of explainable AI in the medical domain.  

As discussed, one of the primary concerns in applying machine learning  to the medial domain is the trust associated with the predictions made by the model. However, completely understanding the function of a classifier, especially a deep learning classifier, is a complex task. Algorithms such as SHAP and Occlusion maps provide insights into how different inputs impact predictions. By analyzing the results, both Shapley values and Occlusion maps reveal that the model is concentrating on the fundamental heart sounds of the PCG wave. Patterns associated with the correct classification of  these signals are evident at those locations or in between them. In fact, this learned insight presents substantial evidence on the medical nature of the PCG signal, and these representations learned by the model seems to have a direct relation to medical procedure followed in digital phonocardiography \cite{Digitacardiograph}. As a result, the behavior of the model appears to be \textit{human-like}. 

The SHAP algorithm can be applied to interpret a model by making global explanations or instance level explanations. However, when the algorithm is applied to the biosignal-based medical domain, the temporal nature of the signal should  also be considered to ensure interperetability of the results. Furthermore, in the medical domain, pre-extracted feature representations such as MFCCs or spectrograms are highly utilized \cite{Clifford2017RecentAnalysis}. As these have a direct relation to the signal (temporal), interpreting them alongside the input signal provides valuable insights into the predictions made by the model.

\bibliographystyle{ieeetr}
\bibliography{bibs/references,bibs/samplenew,bibs/add,bibs/referencesqu}

\end{document}